\documentclass[aps,pre,twocolumn,superscriptaddress,reprint,showpacs,longbibliography]{revtex4-1}

\usepackage{graphicx}
\usepackage{amsmath}
\usepackage[usenames, dvipsnames]{color}
\usepackage[normalem]{ulem}

\newcommand{\kT}{k_{\text{B}}T}
\newcommand{\half}{\ensuremath{\tfrac{1}{2}}}

\newcommand{\Normal}{\mathcal{N}}
\newcommand{\hamil}{\mathcal{H}}
\newcommand{\dt}{\Delta t}
\newcommand{\work}{W}
\newcommand{\heat}{Q}
\newcommand{\shadowwork}{\work_{\rm shad}}			
\newcommand{\protwork}{\work_{\rm prot}}			
\newcommand{\totwork}{\work_{\rm tot}}			
\newcommand{\powerSS}{{\mathcal P}_{\rm ss}}

\newcommand{\rev}{\tilde}
\newcommand{\protocol}{\Lambda }
\newcommand{\F}{\protocol}
\newcommand{\R}{\rev{\F}}
\newcommand{\traj}{X} 
\newcommand{\PP}[2]{P\big[\, #1\, \big|\, #2\,\big]}

\newcommand{\x}{r}	
\newcommand{\wiener}{\, \mathsf W} 
\newcommand{\var}[1] {\mathrm{var}\left( #1 \right)}
\newcommand{\cov}[1] {\mathrm{cov}\left( #1 \right)}
\newcommand{\f}{\frac}

\begin{document}

\title{Using Nonequilibrium Fluctuation Theorems to Understand and Correct Errors in Equilibrium and Nonequilibrium Simulations of Discrete Langevin Dynamics} 

\author{David A.\ Sivak}
\email{david.sivak@ucsf.edu}
\altaffiliation{Current address:  Center for Systems and Synthetic Biology, University of California, San Francisco, California 94158, USA}
\affiliation{Physical Biosciences Division, Lawrence Berkeley National Laboratory, Berkeley, California 94720, USA}

\author{John D.\ Chodera}
\affiliation{Computational Biology Center, Memorial Sloan-Kettering Cancer Center, New York, New York 10065, USA}

\author{Gavin E.\ Crooks}
\affiliation{Physical Biosciences Division, Lawrence Berkeley National Laboratory, Berkeley, California 94720, USA}

\begin{abstract}
Common algorithms for computationally simulating Langevin dynamics must discretize the stochastic differential equations of motion. These resulting finite time step integrators necessarily have several practical issues in common: Microscopic reversibility is violated, the sampled stationary distribution differs from the desired equilibrium distribution, and the work accumulated in nonequilibrium simulations is not directly usable in estimators based on nonequilibrium work theorems. Here, we show that even with a time-independent Hamiltonian, finite time step Langevin integrators can be thought of as a driven, nonequilibrium physical process. Once an appropriate work-like quantity is defined---here called the \emph{shadow work}---recently developed nonequilibrium fluctuation theorems can be used to measure or correct for the errors introduced by the use of finite time steps. In particular, we demonstrate that amending estimators based on nonequilibrium work theorems to include this shadow work removes the time step dependent error from estimates of free energies. We also quantify, for the first time, the magnitude of deviations between the sampled stationary distribution and the desired equilibrium distribution for equilibrium Langevin simulations of solvated systems of varying size. While these deviations can be large, they can be eliminated altogether by Metropolization or greatly diminished by small reductions in the time step. Through this connection with driven processes, further developments in nonequilibrium fluctuation theorems can provide additional analytical tools for dealing with errors in finite time step integrators.
\end{abstract}

\date{\today}
\maketitle

\section{Introduction}
In the computational natural sciences, dynamic properties of a stochastic system are often calculated using simple numerical integrators for Langevin dynamics~\cite{Langevin1908},
\begin{subequations}
\begin{align}
d\x & = v\ dt \\
dv & = \frac{f(t)}{m} \,dt - \gamma v\ dt + \sqrt{ \frac{2\gamma}{\beta m} }\ d\wiener(t), 
\end{align}
\end{subequations}
where the system is driven from equilibrium by a time-dependent Hamiltonian $\hamil(t)$. In the simplest case of a single stochastic particle, $\x$ and $v$ are time-dependent position and velocity, $m$ is mass, $f$ is force, $\beta=1/\kT$, $k_\text{B}$ is Boltzmann's constant, $T$ is the temperature of the environment, $\gamma$ is a friction coefficient (with dimensions of inverse time), and $\wiener(t)$ is a standard Wiener process. The force is determined by the derivative of the potential energy, $f \equiv -\,\partial\hamil/\partial \x$. For multidimensional, multiparticle systems, $\x$, $v$, $f$, and $d\wiener$ are vectors, and $m$ is a diagonal matrix. 

In order to simulate Langevin dynamics on a digital computer, it is necessary to adopt some approximate algorithm that divides time into discrete steps~\cite{Frenkel2002}. However, most such schemes have an inherent problem: Even with a time-independent Hamiltonian, they do not preserve the canonical equilibrium distribution determined by $\hamil$ nor do they satisfy microscopic reversibility. (By reversibility we mean that the probability of sampling a particular trajectory starting from equilibrium is equal to the probability of sampling the trajectory's time reversal, reversing velocities if necessary.) We show that these pathologies arise because discrete time step integrators of Langevin dynamics can be viewed as simulations of artificial driven nonequilibrium dynamics. This perspective has the advantage that the complications generated by this unwanted but inevitable breaking of time-reversal symmetry can be understood and remedied in a controlled and systematic fashion with insights from nonequilibrium statistical thermodynamics~\cite{Athenes2004,Adjanor2005,Adjanor2006,Lechner2006}.

We can appreciate some of the problems inherent in finite time step Langevin dynamics by first considering the zero friction limit, $\gamma=0$, with a time-independent Hamiltonian, where Langevin dynamics reduces to deterministic Newtonian dynamics. A simple, popular integrator for Newtonian dynamics is the velocity Verlet algorithm~\cite{Swope1982, Tuckerman1992},
\begin{subequations}
\label{velocityverlet}
\begin{align}
v(n+\tfrac{1}{2}) &= v(n) + \frac{\dt} {2}\ \frac{f(n)}{m} \\
\x(n+1) &= \x(n) + \dt \ v(n+\tfrac{1}{2}) \\ 
v(n+1) &= v(n+\tfrac{1}{2}) + \frac{\dt}{2}\ \frac{f(n+1)}{m} \ .
\end{align}
\end{subequations}
Because of the finite time step, the trajectories generated by this algorithm are inaccurate: They do not faithfully follow the precepts of Newtonian mechanics. Also, the actual energy of the system is not conserved, but rather it fluctuates from one time step to the next. However, the velocity Verlet integration scheme is symplectic (in that the Jacobian of the transformation from old to new positions and velocities is unity, and therefore the phase-space volume is conserved~\cite{SanzSerna:1992ut}), which ameliorates some problems due to the finite time step. For example, although a finite time step symplectic integrator does not conserve the energy of the system Hamiltonian, it does conserve the energy of a shadow Hamiltonian, which is close to the desired Hamiltonian if the time step is not too large~\cite{Yoshida1990,Frenkel2002}. For sufficiently small timesteps, this conservation of the shadow Hamiltonian prevents long-term drift in the system Hamiltonian over the duration of the simulation.  

Essentially, a finite time step dynamics performs work on the system, over-and-above any work due to intentional perturbations from a time-dependent Hamiltonian~\cite{Lechner2006}. We can imagine this finite time step integration scheme in the following way. At the beginning of each time step, we first perturb the system Hamiltonian such that it becomes the shadow Hamiltonian, changing the energy of the system. The symplectic integrator then updates the position and velocity [Eq.~\eqref{velocityverlet}], perfectly preserving the shadow energy of the shadow Hamiltonian. We then switch the Hamiltonian back to the original one, again perturbing the energy. The net change in the energy of the system during this time step is due to work performed on the system by perturbing back and forth between the system and shadow Hamiltonian. We can determine this shadow work (also known as \emph{error work}~\cite{Lechner2006} or an \emph{effective energy} change~\cite{Bussi2007}) during each time step by measuring the difference in energy using the system Hamiltonian, so we do not need to know the form of the shadow Hamiltonian. This shadow work is distinct from any protocol work applied to the system due to explicit, time-dependent perturbations of the system Hamiltonian. Note that Markov-chain Monte Carlo (MCMC) simulations do not generate shadow work~\cite{Lelievre2010a} because the dynamics satisfies detailed balance explicitly, which ensures that the trajectories are microscopically reversible~\cite{Tolman1938} and that the appropriate equilibrium ensemble is preserved for a time-independent Hamiltonian~\cite{Frenkel2002}. 

Discretizations of continuous-time Langevin dynamics are essentially a combination of deterministic and stochastic dynamics, and, as a result, they suffer from a combination of problems. With a finite time step, the deterministic parts of the dynamics tend to pump energy into the system in the form of shadow work, driving the system away from equilibrium, whereas the stochastic parts of the dynamics relax the velocities back toward the equilibrium Maxwell-Boltzmann distribution, removing energy from the system in the form of heat. It follows that, even for a system with a Hamiltonian that is \emph{explicitly} time-independent, a finite-time-step Langevin dynamics has an \emph{effective} Hamiltonian alternating between the system Hamiltonian and the shadow Hamiltonian, and thus actually simulates a driven, nonequilibrium system, with a net energy flow. Microscopic time-reversal symmetry is broken, and in general we can not determine the steady-state, nonequilibrium distribution. These difficulties may be circumvented by reducing the time step, but at the cost of increasing the computational effort required to simulate the same interval of time; this hardly constitutes a satisfactory resolution. 

The main point of this paper is this interpretation of the errors induced by discrete simulation of Langevin dynamics, in terms of a driven thermodynamic process. This perspective forms a bridge between the study of numerical integrators and the rapidly expanding field of nonequilibrium statistical mechanics, permitting the invocation of a wide array of nonequilibrium work fluctuation relations to characterize and correct for biases in estimates of equilibrium and nonequilibrium thermodynamic quantities.

\section{Concrete Integrator}
\begin{figure}[t]
\begin{center}
\includegraphics{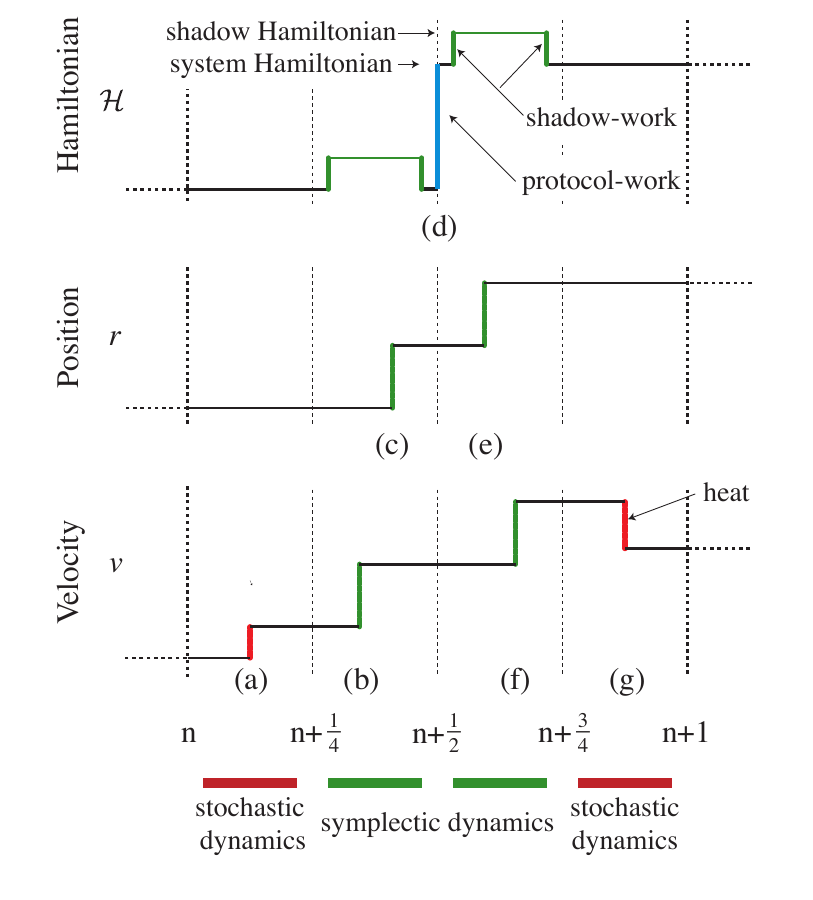} 
\caption{
Timeline for the Langevin integrator~\eqref{BusPar}. The part labels (3a-3g) correspond to the substeps of Eq.~\eqref{BusPar}. The stochastic substep \eqref{heat1} randomizes the velocity, transferring heat between the system and environment, while the Hamiltonian is fixed and the position unchanged. We then switch from the system to the shadow Hamiltonian, performing shadow work on the system. Substeps \eqref{dV1} and \eqref{dX1} update the velocity and then the position according to the symplectic dynamics of the shadow Hamiltonian, exactly conserving the energy. We next switch back to the system Hamiltonian (performing shadow work), and in \eqref{dH} update the system Hamiltonian~from $\hamil(n)$ to $\hamil(n+1)$, according to the prescribed protocol $\Lambda$. This action performs protocol work on the system. We switch back to the shadow Hamiltonian (doing shadow work), symplectically update position and then velocity (\ref{dX2},\ref{dV2}), and then restore the system Hamiltonian (again performing shadow work). Finally, we conclude with another velocity-randomization substep \eqref{heat2}.}
\label{VVVRfig}
\end{center}
\end{figure}

We demonstrate the utility of this perspective for an integration scheme that is explicitly time-symmetric, that cleanly separates the stochastic and deterministic parts of the dynamics, and for which the deterministic parts are symplectic and the stochastic parts are detailed balanced. This construction allows a clean separation of the system's energy change into work, shadow work, and heat, simplifying our analysis in terms of a driven nonequilibrium process. Fortunately, integrators with these properties have received recent attention~\cite{Adjanor2005,Adjanor2006,Bussi2009,Bou-Rabee2010,Lelievre:2012tg}. As a concrete example, we consider the integrator used by Bussi and Parrinello~\cite{Bussi2007}, where we make the Hamiltonian update explicit:
\begin{subequations}
\label{BusPar} 
\begin{align}
v(n+\tfrac{1}{4}) & = \sqrt{a}\ v(n) + \sqrt{\f{1\text{-}a}{\beta m}}\ \Normal^+(n) \label{heat1} \\
v(n+\tfrac{1}{2}) & = v(n+\tfrac{1}{4}) + \frac{\dt} {2}\ \frac{f(n)}{m} \label{dV1} \\
\x(n+\tfrac{1}{2}) &= \x(n) + \frac{\dt}{2} \, v(n+\tfrac{1}{2}) \label{dX1} \\
\hamil(n) & \rightarrow \hamil(n+1) \label{dH} \\
\x(n+1) &= \x(n+\tfrac{1}{2}) + \frac{\dt}{2}\, v(n+\tfrac{1}{2}) \label{dX2} \\ 
v(n+\tfrac{3}{4}) & = v(n+\tfrac{1}{2}) + \frac{\dt}{2}\ \frac{f(n+1)}{m} \label{dV2} \\
v(n+1) & = \sqrt{a} \ v(n+\tfrac{3}{4}) + \sqrt{\f{1\text{-}a}{\beta m}}\ \Normal^-(n+1) \label{heat2}
\end{align}
\end{subequations}
Here, $\dt$ is the time step by which the simulation clock is advanced, $f(n)$ is the force at position $r(n)$ due to the Hamiltonian $\hamil(n)$, $a=\exp({- \gamma\, \dt})$, and $\Normal^+$ and $\Normal^-$ are independent, normally distributed random variables with zero mean and unit variance (hence, when scaled by $(\beta m)^{-1/2}$, distributed according to the equilibrium Maxwell-Boltzmann velocity distribution). The first and last substeps~(\ref{heat1},\ref{heat2}) are stochastic, Markovian, and detailed-balanced (with respect to the canonical measure) velocity randomizations, which leave the position unchanged. The five middle substeps~(\ref{dV1}-\ref{dV2}) constitute the deterministic velocity Verlet integrator \eqref{velocityverlet}, with the midpoint Hamiltonian update made explicit. The order of substeps and the effective Hamiltonian switches are illustrated in Fig.~\ref{VVVRfig}. Note that the deterministic substeps (\ref{dV1},\ref{dX1},\ref{dX2},\ref{dV2}) are each individually symplectic.

\section{Nonequilibrium thermodynamics}

A central relation of driven, nonequilibrium thermodynamics~\cite{Bustamante2005,Cleuren2007,Jarzynski2011,Spinney2012} relates the microscopic irreversibility of trajectories to the work $\work[\traj,\Lambda]$ performed on the system during the forward protocol~\cite{Crooks1998,Crooks1999a,Crooks1999c}:
\begin{align}
\ln \frac{\PP{\traj}{\Lambda} }{ \PP{\rev\traj}{\rev\Lambda}} &= { \beta\work[\traj,\Lambda] } - \beta\Delta F_{\rm eq}[\Lambda] \label{microirr} \ . 
\end{align}
Here, $\traj$ is a trajectory through phase space between time $0$ and $N\dt$, $\Lambda$ represents a protocol for perturbing the system (typically through the time dependence of the system Hamiltonian), $\Delta F_{\rm eq}[\Lambda]$ is the free energy difference between the equilibrium distributions for the initial and final values of the system Hamiltonian, and $\PP{\traj}{\Lambda}$ is the probability of the trajectory, given the protocol and an initial equilibrium ensemble. The time-reversed protocol $\rev\Lambda$ (time-reversed trajectory $\rev\traj$) retraces the same series of perturbations (phase-space transitions) as the forward protocol $\Lambda$ (forward trajectory $\traj$), but under time inversion and hence in reverse. Subject to a protocol, a driven system is microscopically reversible if the probability of a trajectory and its time reversal are identical, and therefore the work imposed by the protocol equals the free energy change~\cite{Crooks2011b}.

It is straightforward to extend this fluctuation theorem to mixed stochastic-deterministic dynamics, such as the Langevin integrator, Eq.~\eqref{BusPar}, provided that the individual substeps satisfy this symmetry. It is for this reason that we insist on a clean separation of the deterministic and stochastic substeps.

The total work $W = \sum_n W^{(n)}$ is the sum of the contributions $W^{(n)}$ from individual steps. The total change in energy $\Delta E$ during the step $n \rightarrow n+1$ can be cleanly separated into heat $\heat$, protocol work $\protwork$, and shadow work $\shadowwork$:
\begin{subequations}
\begin{align}
\Delta E &= \heat + \work \label{firstlaw} \\ 
\notag &= \heat + \protwork + \shadowwork \\
\heat &= \Delta E_a + \Delta E_g \label{defineHeat} \\
\protwork &= \Delta E_d \label{defineProtWork} \\
\shadowwork &= \Delta E_b + \Delta E_c + \Delta E_e + \Delta E_f \label{definePseudoWork} \ .
\end{align}
\end{subequations}
Here, $\Delta E_{a\text{-}g}$ are the energy changes during the corresponding substeps of Eq.~\eqref{BusPar}. Heat is the energy exchanged with the thermal environment, protocol work is the energy change due to deliberate manipulation of the Hamiltonian (i.e., the explicit time-dependence of the system Hamiltonian), and shadow work is the energy change due to alternation between the system and shadow Hamiltonians, resulting from the finite time step of the symplectic part of the integrator. The essential distinction between heat and work is that heat flow is change of the system energy due to change in the \emph{current} distribution over microstates, whereas work is change of energy due to change in the \emph{equilibrium} distribution over microstates.

The stochastic velocity randomization substeps obey Eq.~\eqref{microirr} since they are balanced, in that they preserve the canonical equilibrium distribution~\cite{Crooks1998}. The set of deterministic velocity Verlet substeps also obeys Eq.~\eqref{microirr}, so long as the total work includes the shadow work~\cite{Athenes2004,Lechner2006}, since the dynamics is symplectic and microscopically reversible with respect to the shadow Hamiltonian~\cite{Yoshida1990,Frenkel2002}. Since both the deterministic and stochastic substeps are Markovian, it follows that we can safely intermix the two dynamics, and \eqref{microirr} still holds. 

It therefore follows that the Langevin integrator obeys various derived relations of nonequilibrium statistical dynamics, such as the Jarzynski equality~\cite{Jarzynski1997a}, fluctuation relations~\cite{Evans1994,Crooks1998}, interrelations between path ensemble averages~\cite{Crooks1999a, Hummer2001a} and various interrelations between dissipation and time asymmetry~\cite{Gaspard2004a,Jarzynski2006a,Kawai2007a,Feng2008}. Furthermore, by its separation of protocol work and shadow work, the Langevin integrator permits the separation of the respective contributions to microscopic irreversibility of deliberate perturbation (physically meaningful) and the finite time step (a discretization artifact). Notably, the statistics of the protocol work alone systematically deviate from those of the total work, and hence lead to biased inference when using the machinery of nonequilibrium thermodynamics. In Secs.~\ref{sec:eqStat} and \ref{sec:nonEqStat} we explicitly demonstrate this underappreciated point.

\section{``Equilibrium'' simulations sample perturbed distributions}
It is common practice in the study of the equilibrium properties of molecular systems to use a single finite-time-step mixed stochastic and deterministic dynamical simulation to sample from an equilibrium distribution. However, this distribution departs from the true equilibrium distribution for the system Hamiltonian, a distribution that we can now understand as the steady state due to driving by the finite time step. Thus a question of significant practical interest presents itself: How far from equilibrium is the effective nonequilibrium steady state induced by this time discretization for a system with a time-independent Hamiltonian? Since the explicit system Hamiltonian is unchanging, no protocol work is performed, and thus our analysis in this section focuses on the shadow work alone. Practitioners commonly estimate artifactual errors by monitoring some essentially arbitrary, yet easily measured, observable of the system, such as the total energy. However, we can exploit recent advances in nonequilibrium statistical dynamics to provide a principled characterization of how far the system is driven from equilibrium~\cite{Sivak:2012ab}.

The natural measure of this instantaneous distance that the system has been driven away from equilibrium is the difference between a nonequilibrium free energy~\cite{shawFaucet,Gaveau:1997ul} $F_{\rm neq} \equiv \langle E\rangle - T S$ and the corresponding equilibrium free energy $F_{\rm eq}$ for the given Hamiltonian $\hamil$. If the Hamiltonian were held constant and the (previously driven) system were allowed to relax to equilibrium, this deviation from the equilibrium free energy would represent the heat that would be lost to the environment, or equivalently the maximum work that could be imparted to a mechanically coupled system. For the perturbations imposed by the discrete dynamics, this nonequilibrium free-energy deviation is approximated near equilibrium by~\cite{Sivak:2012ab}
\begin{align}
\Delta F_{\rm neq} \equiv F_{\rm neq} - F_{\rm eq} \approx \half \big[ \langle \shadowwork \rangle - (t_{\rm f}-t_{\rm i})\powerSS \big] \label{equ:FneqApp}\ ,
\end{align}
where $\shadowwork$ is the shadow work over the whole simulation, $\powerSS$ is the power (work per unit of time) once transients have died off and the system has settled into a nonequilibrium steady state, and $t_{\rm f}-t_{\rm i}$ is the total simulation time. Normalizing this nonequilibrium free-energy deviation by the size of the system (number of degrees of freedom) provides a natural measure of how far from equilibrium each degree of freedom is on average.

To estimate the nonequilibrium steady-state free-energy deviation for a molecular system, we simulate cubic boxes of TIP3P waters of various sizes, both with and without constraints on the water O-H and H-H interatomic distances. (See the Appendix for simulation details.) Initial coordinates and momenta are sampled from equilibrium in an isothermal-isobaric (NPT) ensemble (that is, an ensemble that maintains constant number of particles, constant pressure, and constant temperature) at 1~atm and 298~K using the generalized hybrid Monte Carlo (GHMC) integrator~\cite{Lelievre:2012tg,Horowitz1991}. These initial conditions are simulated for $M$ steps with the Langevin integrator [Eq.~\eqref{BusPar}] at constant volume (using a collision rate $\gamma = 9.1$/ps) to measure the nonequilibrium work to reach steady state, followed by an additional $M$ steps to measure the steady-state power. We have determined that, for all systems and time steps simulated, $M = 1028$ steps is sufficient to reach steady state (see Fig.~\ref{fig:converge}). We have also calculated the statistical uncertainty according to Eq.~\eqref{equ:waterUncertain}.

\begin{figure}[t]
\begin{center}
\includegraphics[width=1.0\columnwidth]{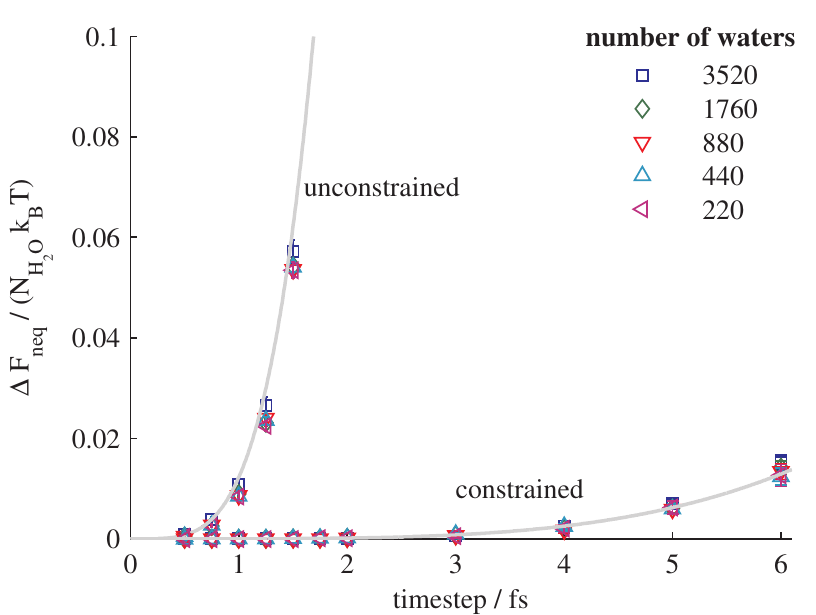}
\caption{Nonequilibrium free-energy deviation for boxes of TIP3P water, normalized by number of waters. Nonequilibrium free energies for various system sizes (220 to 3520 TIP3P waters) are shown for both unconstrained (left curve) and constrained (right curve) simulations, normalized by the number $N_{\rm H_2O}$ of waters in the system and the thermal energy $\kT$. Error bars show 95\% confidence intervals. Gray curves show empirical fits of the form $a \cdot \Delta t^4$, with $a = 1.23 \cdot 10^{-2} \text{fs}^{-4}$ for unconstrained simulations and $a = 9.97 \cdot 10^{-6} \text{fs}^{-4}$ for constrained simulations.}
\label{fig:waterBox}
\end{center}
\end{figure}

Because the system (a periodic water box) is homogeneous, it is possible to collapse all system sizes onto universal curves describing the nonequilibrium free-energy deviation per molecule as a function of time step for unconstrained and constrained systems, respectively (Fig.~\ref{fig:waterBox}). For the unconstrained system, whose numerical integration becomes unstable beyond $\Delta t = 1.5$~fs, the nonequilibrium free-energy deviation $\Delta F_{\rm neq}$ rapidly rises as the time step surpasses the typical time step employed for flexible systems, $\Delta t \approx 1$~fs. For a system of 220 waters, for example, $\Delta F_\mathrm{neq} = 11.4 \ \pm\ 0.2~\kT$ at $\Delta t = 1$~fs. For constrained water boxes, however, $\Delta F_\mathrm{neq}$ reaches this magnitude only at large time steps---here, $\Delta t \approx 5$~fs, not far from the stability limit at 6~fs and well beyond $2$~fs, the standard time step for biomolecular simulations.

Empirically, the nonequilibrium free-energy deviation $\Delta F_\mathrm{neq}$ for both unconstrained and constrained systems appears to show a quartic dependence on the time step $\Delta t$ (Fig.~\ref{fig:waterBox}, gray curves), such that 
\begin{align}
\frac{ \Delta F_\mathrm{neq} }{N_{\rm H_2O} \kT} &\approx a \cdot \Delta t^4 \ ,
\end{align} 
where the prefactor $a$ depends strongly on whether constraints are employed; see the caption of Fig.~\ref{fig:waterBox}. This trend is consistent with earlier work observing the strong dependence of Metropolization acceptance probabilities on time step~\cite{Beskos:arxiv} and highlights how small reductions in time step can rapidly reduce the deviations of the sampled steady-state distribution from the desired equilibrium distribution defined by the system Hamiltonian $p_{\rm eq}(x) \propto \exp[-\beta \mathcal{H}(x)]$, without unduly burdensome computational cost. We detail in Sec.~\ref{sec:eqStat} some methods that correct for these nonequilibrium perturbations. Even in the absence of correction procedures, the above calculation represents a thermodynamically meaningful determination of the deviation from the desired equilibrium sampling associated with the continuous Langevin equation of motion, as a function of simulation parameters.

\section{Multivariate Fluctuation Theorem}
\label{sec:multiFT}
We seek an analytical framework that describes the correlation between the shadow work (performed by integration) and the protocol work (due to explicit Hamiltonian changes). We want this framework to provide a generic method to characterize the effect that shadow work has on the distribution of protocol work, and specifically on the time-reversal symmetry [Eq.~\eqref{microirr}] that protocol work would satisfy in its absence. Furthermore, we want this framework to suggest systematic techniques to correct for these distorting effects. We propose such a framework through the generalization of work fluctuation theorems to the context of two sources of work. These results, although formulated specifically for our situation of explicit and artifactual work, are entirely general to situations involving any two sources of work.

Rearrangement of Eq.~\eqref{microirr} and splitting the work into two distinct work contributions $W_1,W_2$ gives
\begin{equation}
\PP{\traj}{\F} = \PP{\rev\traj}{\R} e^{\beta\left\{ \work_1[\traj,\F] + \work_2[\traj,\F] - \Delta F_{\rm eq}[\F] \right\} } \ . 
\end{equation}
Multiplication by delta functions of the two works, $\delta(\work_1[\traj,\F]-\protwork)\delta(\work_2[\traj,\F]-\shadowwork)$, and integration over all trajectories produces what we refer to as the multivariate fluctuation theorem,
\begin{align}
\label{equ:multiFT}
\f{ P_{\F}(\protwork,\shadowwork) }{ P_{\R}(-\protwork,-\shadowwork) }  = e^{\beta(\protwork+\shadowwork-\Delta F_{\rm eq})} \ .
\end{align}
This is a special case of the generalized detailed fluctuation theorem for joint probabilities of Garc{\`\i}a-Garc{\`\i}a, et al.~\cite{GarciaGarcia:2010iv,GarciaGarcia:2012fo}. Equation~\eqref{equ:multiFT} gives an expression in terms of the excess work $\protwork+\shadowwork-\Delta F_{\rm eq}$ for the ratio of the joint probability distributions over protocol and shadow works realized during the forward and reverse protocols, respectively.

Equation~\eqref{equ:multiFT} can be trivially extended to arbitrary decompositions of the total work, where each component corresponds to a group of individual work steps. It thus represents a generalization of the work fluctuation theorem~\cite{Crooks1999a} to contexts with multiple sources of work. From Eq.~\eqref{equ:multiFT}, several other modified fluctuation theorems can be derived that modify a standard fluctuation theorem for one of the works with an exponential average over the other work. For example, in Sec.~\ref{sec:eqStat}, we derive a Jarzynski equation modified by the presence of shadow work [Eq.~\eqref{equ:modJar}], and, in Sec.~\ref{sec:nonEqStat}, we derive a similarly modified integrated transient fluctuation theorem [Eq.~\eqref{equ:ITFT2}].

\section{Recovering equilibrium statistics from nonequilibrium simulations\label{sec:eqStat}}
Now that we are equipped with our new interpretation of finite-time-step Langevin dynamics as a driven nonequilibrium process even in the absence of an explicit driving force, nonequilibrium thermodynamics affords various approaches for recovering true equilibrium properties of the system. 

One approach is to maintain the simulation at equilibrium by incorporating Monte Carlo moves that conditionally accept or reject candidate trajectory segments or single time steps, for example by using the Metropolis criterion $P_{\text{accept}} = \text{min}(1, \exp\{-\beta \shadowwork\})$~\cite{Metropolis1953}. In order to maintain detailed balance, the velocity must be inverted if the proposed state is rejected~\cite{Lelievre2010a}, which may lead to increased correlation times. Applied to single time steps, this is essentially the idea behind the GHMC integrator~\cite{Horowitz1991, Lelievre2010a}, and when applied to trajectory segments, this is the idea behind work-bias Monte Carlo~\cite{Athenes2002} and nonequilibrium candidate Monte Carlo~\cite{Nilmeier:2011gr} simulations. In either case, Metropolization results in an MCMC process that samples the true equilibrium distribution.

Another approach to recovering equilibrium statistics is to perform a Monte Carlo sampling of trajectories~\cite{Sun2003, Atilgan2004}, generating an ensemble of trajectories weighted by the Boltzmann-weighted work over the entire trajectory, $\exp\{-\beta \shadowwork\}$. This approach allows both accurate equilibrium statistics and realistic dynamics, albeit at a potentially high computational cost. 

Instead of sampling equilibrium trajectories, we can alternatively apply nonequilibrium relations, such as the Jarzynski equality~\cite{Jarzynski1997a} and path ensemble averages~\cite{Crooks1999a, Hummer2001a, Minh2009, Minh2011}, to directly recover equilibrium properties from the statistics of a driven system, essentially by reweighting trajectories by $\exp\{-\beta \work_{\rm tot}\}$, where it is important that the work includes both the protocol work and the shadow work. Note that the initial configurations must be sampled from the correct equilibrium ensemble, which can be accomplished with a standard MCMC process, or with one of the approaches discussed above, such as GHMC simulation. 

\begin{figure*}[t]
\begin{center}
\includegraphics{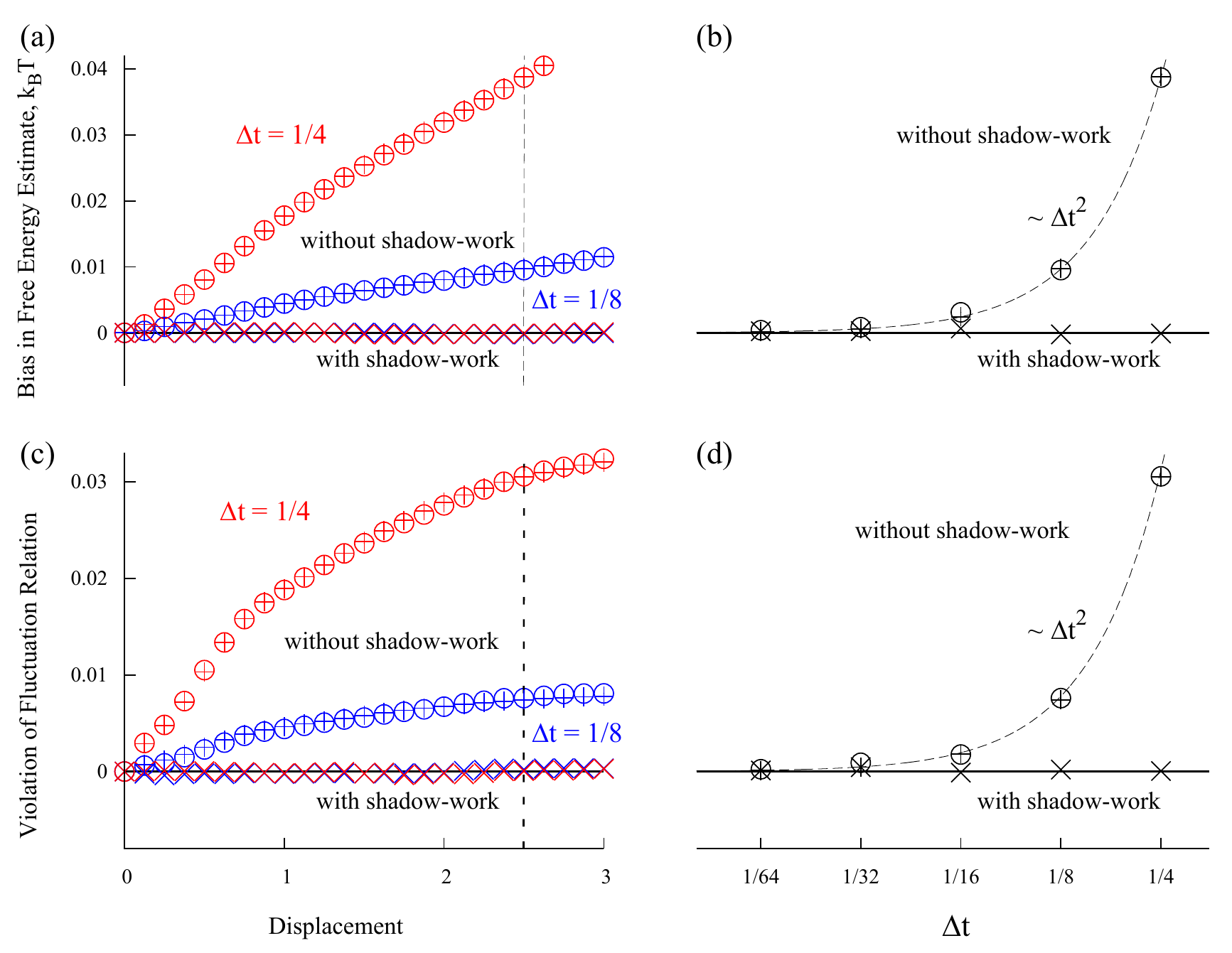}
\caption{
Ignoring shadow work in Langevin simulations leads to systematic errors in inference of both equilibrium and nonequilibrium statistics. Results are from Langevin simulations of $10^8$ independent realizations of a quartic potential $U = \tfrac{1}{4}(x-x_{\text min})^4$ uniformly translating with velocity $1/2$, starting from equilibrium, with unit temperature, mass, spring constant, and friction coefficient. Standard errors are smaller than symbol size. 
(a) Error in free energy calculated from the Jarzynski equality~\eqref{equ:Jarzynski} as a function of position of the quartic potential, neglecting shadow work (circles) and including shadow work ($\times$s), for $\dt$ of 1/4 (red) and 1/8 (blue). The exact free energy change is zero. The error in the naive Jarzynski estimator (circles) is entirely captured by the correction term $- \ln \langle e^{-\beta \shadowwork} \rangle_{\R}$ ($+$ signs) from Eq.~\eqref{equ:modJar}, as can be seen by the agreement of these symbols to within statistical error.
(b) Semilog plot of error in the Jarzynski free energy estimate after the quartic potential has moved to $\x=2.5$, as a function of time step length, neglecting shadow work (circles) and including shadow work~($\times$s). Also shown is the correction term from Eq.~\eqref{equ:modJar} ($+$ signs).
(c) Ratio of left-hand side and right-hand side of the integrated transient fluctuation theorem (ITFT) as a function of position of the quartic potential, neglecting shadow work [circles, Eq.~\eqref{equ:ITFT2}] and including shadow work [$\times$s, Eq.~\eqref{equ:ITFT}], for $\dt$ of 1/4 (red) and 1/8 (blue). The error in the naive ITFT ratio is entirely captured by the correction factor $\left\langle e^{-\beta \totwork} \right\rangle_{\protwork>0} / \left\langle e^{-\beta \protwork} \right\rangle_{\protwork>0}$ ($+$ signs), as can be seen by the agreement of these symbols to within statistical error.
(d) Semilog plot of the ITFT ratio after the quartic potential has moved to $\x=2.5$, as a function of time step length $\dt$, neglecting shadow work (circles) and including shadow work~($\times$s). Also shown is the correction factor $\left\langle e^{-\beta \totwork} \right\rangle_{\protwork>0} / \left\langle e^{-\beta \protwork} \right\rangle_{\protwork>0}$ ($+$ signs).
}
\label{fig:jarITFTFig}
\end{center}
\end{figure*}

We now demonstrate the importance of including the shadow work by using the Jarzynski equality to estimate free energy changes in a simple model system. The Jarzynski equality~\cite{Jarzynski1997a} relates the equilibrium free energy change, resulting from some perturbation of the system, to the exponential average of the work incurred during many realizations of the system response to that perturbation,
\begin{subequations}
\begin{align}
\label{equ:Jarzynski}
\beta \Delta F_{\rm eq} &= - \ln \left\langle e^{-\beta \work}\right\rangle_{\Lambda} \\
&= - \ln \left\langle e^{-\beta(\protwork + \shadowwork) }\right\rangle_{\Lambda} \ . \label{equ:jarWork}
\end{align}
\end{subequations}
In the second line, we have explicitly split the effective thermodynamic work into protocol and shadow work. Here, angled brackets with subscript $\Lambda$ indicate expectations over trajectories starting in the equilibrium distribution for the initial value of the Hamiltonian $\hamil(0)$ and integrated according to Eq.~\eqref{BusPar}, with the Hamiltonian evolving according to $\Lambda$. Although standard Langevin integrators are used in myriad multidimensional contexts, we examine in Fig.~\ref{fig:jarITFTFig} the shadow work contribution in a simple one-dimensional system to suggest the ubiquity of the issues raised here. In particular, we consider a particle in thermal contact with the environment, subject to a quartic potential that is initially stationary and then translated at a constant velocity. The exact free energy change is zero. When one uses only the protocol work (neglecting the shadow work), the Jarzynski free energy estimate empirically shows a systematic error that scales roughly as $\dt^2$ [Figs.~\ref{fig:jarITFTFig}a,b, circles]. Using the total thermodynamic work (including the shadow work) eliminates this error, and the Jarzynski estimator gives the correct free energy change [Figs.~\ref{fig:jarITFTFig}a,b, $\times$s]. In Fig.~\ref{fig:jarITFTFig}, standard errors are calculated from $10^8$ independent simulations and are smaller than the symbol size. The $y$ axis is the same in the left and right sub-figures. 

We can understand the origin of this error by analyzing our estimator in terms of the multivariate fluctuation theorem [Eq.~\eqref{equ:multiFT}] derived above in Sec.~\ref{sec:multiFT}. Rearranging Eq.~\eqref{equ:multiFT}, decomposing the joint probability into the marginal and conditional probabilities, 
\begin{equation}
P_{\R}(-\protwork,-\shadowwork) = P_{\R}(-\protwork)P_{\R}(-\shadowwork|-\protwork) \ ,
\end{equation}
and integrating over the shadow work, we find that when ignoring the contributions of shadow work, the Jarzynski estimator of the free energy $\beta \widehat{\Delta F}_{\rm eq} \equiv -\ln\left\langle e^{-\beta\protwork}\right\rangle_{\Lambda}$ has a systematic bias from the true free energy change $\beta\Delta F_{\rm eq}$ that is a function of the distribution of shadow works: 
\begin{equation}
\beta \widehat{\Delta F}_{\rm eq} = \beta\Delta F_{\rm eq} - \ln \langle e^{-\beta \shadowwork} \rangle_{\R} \ . \label{equ:modJar}
\end{equation}
Empirically, the correction term $- \ln \langle e^{-\beta \shadowwork} \rangle_{\R}$ [Figs.~\ref{fig:jarITFTFig}a,b, $+$ signs] reproduces the error in the Jarzynski estimator without shadow work, $\beta \widehat{\Delta F}_{\rm eq}$.

The correction factor $\gamma \equivÊ\langle e^{-\beta \shadowwork} \rangle_{\R}$ is analogous to the correction factor that appears in the Jarzynski equality with feedback~\cite{Sagawa:2010df}. Curiously, the correction to the Jarzynski estimator is solely a function of the shadow work distribution, and, in particular, does not explicitly depend on correlations between the shadow work and the protocol work.

\vspace{5ex}
\section{Correcting nonequilibrium fluctuation theorems\label{sec:nonEqStat}}
In addition to these errors for equilibrium estimators during simulations with an explicitly time-independent Hamiltonian, ignoring the contribution of shadow work leads to systematic errors in estimates of \emph{nonequilibrium} quantities when the Hamiltonian is explicitly time dependent: The simulated system is actually subject to a different Hamiltonian than the system one, and thus the probability distribution of protocol works does not obey the relevant time-reversal symmetry~\eqref{microirr}. We quantitate this time-reversal asymmetry by examining violations of the integrated transient fluctuation theorem (ITFT)~\cite{Ayton:2001wf}, which for time-symmetric protocols relates the ratio of the probabilities of realizing a negative and a positive total work, respectively, to the exponentially-weighted total work, conditional on the total work being positive: 
\begin{equation}
\label{equ:ITFT}
\frac{P(\totwork<0)}{P(\totwork>0)} = \left\langle e^{-\beta \totwork} \right\rangle_{\totwork>0} \ . 
\end{equation}
This relation follows directly from Eq.~\eqref{microirr}.

Manipulating Eq.~\eqref{equ:multiFT} to a similar form produces
\begin{equation}
\frac{P(\protwork<0)}{P(\protwork>0)} = \left\langle e^{-\beta \totwork} \right\rangle_{\protwork>0} \ .
\end{equation}
For this relation to hold, the work in the exponential must be the \emph{total} work, not the protocol work that appears elsewhere in the equation. When one ignores the shadow work and measures only the protocol work, the ratio of the left-hand side and right-hand side, 
\begin{equation}
\label{equ:ITFT2}
\frac{P(\protwork<0)}{P(\protwork>0)} \ \Big/ \left\langle e^{-\beta \protwork} \right\rangle_{\protwork>0} \ ,
\end{equation}
departs from unity to the extent that the protocol work fluctuations do not obey the relevant time-reversal symmetry that the total work fluctuations do.
Departure from unity in Eq.~\eqref{equ:ITFT2} quantifies the violation of the nonequilibrium time-reversal symmetry obeyed by a proper thermodynamic work encompassing all energy changes not related to heat.

Figures~\ref{fig:jarITFTFig}c,d show that, for the simple system described in Sec.~\ref{sec:eqStat}, the protocol work alone (circles) does not obey the nonequilibrium fluctuation relation required of a thermodynamic work (with an error that empirically scales with the square of the time step), but the sum of the protocol and shadow works ($\times$s) does obey it.
The correction factor $\left\langle e^{-\beta \totwork} \right\rangle_{\protwork>0} / \left\langle e^{-\beta \protwork} \right\rangle_{\protwork>0}$ ($+$ signs) reproduces the error in the ITFT ratio neglecting shadow work. Thus, ignoring the shadow work and using the protocol work rather than the total work produces systematic biases in estimators of nonequilibrium quantities (such as the nonequilibrium free energy~\cite{Sivak:2012ab} or the nonequilibrium energetic efficiency~\cite{Sivak:2012gr}).

\section{Epilogue}
For Hamiltonian dynamics, a finite-time-step symplectic integrator conserves a shadow Hamiltonian and is microscopically reversible. But, as we have seen, for Langevin dynamics, discretization of the dynamics leads (even for a time-independent Hamiltonian) to a mixed deterministic-stochastic nonequilibrium dynamics, which preserves the equilibrium distribution of neither the system nor the shadow Hamiltonian and which is not time-reversal symmetric. However, we can measure the work, heat, and shadow work, and thereby separate the respective contributions to time-reversal symmetry breaking of the finite time step and deliberate perturbation. This procedure allows us to apply results from nonequilibrium thermodynamics to characterize in a thermodynamically meaningful way the error produced by finite-time-step integration and to correct for such errors to recover equilibrium and nonequilibrium properties of the system.

While we focus in this paper on work distributions, we note that discrete integrators can also introduce artifacts into other aspects of a system's dynamical evolution, for example, producing erroneous free-particle diffusion coefficients and uniform force-field terminal drifts. These artifacts can be mitigated through time-step rescaling, as discussed in Ref.~\cite{newIntegrator}. Where measurements of work and heat are not required, correct statistics of nonequilibrium trajectories through phase space can be recovered using the Metropolis-adjusted geometric Langevin algorithm of Bou-Rabee and Vanden-Eijnden, which under reasonable conditions on the potential energy is pathwise convergent to the distribution of trajectories for the continuous equations of motion~\cite{BouRabee:2010tb}.

\section*{Acknowledgments}
The authors thank the anonymous referees for suggestions that substantially improved the manuscript.  The authors also thank Manuel Ath\`enes (Commissariat \`a l'E\'nergie Atomique/Saclay), Gabriel Stoltz (CERMICS, Ecole des Ponts ParisTech), Beno\^{i}t Roux (University of Chicago), Jerome P.\ Nilmeier (Lawrence Livermore National Laboratory), Todd Gingrich (UC Berkeley), Jes\'us A.\ Izaguirre (University of Notre Dame), Benedict Leimkuhler (University of Edinburgh), Jason Wagoner (Stanford University), Huafeng Xu and Cristian Predescu (D.\ E.\ Shaw Research) for enlightening discussions and constructive feedback on the manuscript, and Avery A.\ Brooks for help with the illustrations. The authors are grateful to Peter Eastman and Vijay Pande (Stanford University) for their assistance with the OpenMM molecular simulation library. J.\ D.\ C. was supported through a Distinguished Postdoctoral Fellowship from the California Institute for Quantitative Biosciences (QB3) at the University of California, Berkeley. D.\ A.\ S. and G.\ E.\ C. were funded by the Office of Basic Energy Sciences of the U.S. Department of Energy under Contract No. DE-AC02-05CH11231. Water simulations were carried out on the NCSA Forge supercomputer through an allocation (TG-MCB100015) from the Extreme Science and Engineering Discovery Environment (XSEDE), which is supported by National Science Foundation Grant No. OCI-1053575.

\appendix*

\section{Simulation details}
We carried out simulations using the OpenMM GPU-accelerated molecular simulation toolkit~\cite{OpenMM,OpenMM2} (development revision r3314). Cubic water boxes of various sizes (220, 440, 880, 1760, and 3520 waters) were created using the OpenMM Modeller tool and parametrized with TIP3P water~\cite{TIP3P} using the OpenMM Forcefield tool. In constrained simulations, we used the analytical SETTLE algorithm~\cite{SETTLE} to enforce constraints on water O-H and H-H interatomic distances. This Langevin integrator maintains second-order accuracy~\cite{Lelievre:2012tg} when constrained by the RATTLE algorithm~\cite{ANDERSEN:1983wg}, which should produce results identical (to within machine precision) to SETTLE. We truncated Lennard-Jones interactions at 9~\AA and added an analytical long-range dispersion correction~\cite{Shirts:2007db} to account for interactions beyond this cutoff. We handled electrostatics using the reaction-field algorithm~\cite{reaction-field} with an identical cutoff using an exterior dielectric of 78.5.

We sampled initial configurations and momenta from an equilibrium NPT ensemble at 1 atm and 298~K with the GHMC algorithm~\cite{Lelievre:2012tg,Horowitz1991} using a 0.5~fs time step. We controlled pressure using a Monte Carlo molecular-scaling barostat with a proposal size automatically determined during equilibration~\cite{barostat,barostat2}. After initial equilibration for 250 000 steps, we sampled configurations and momenta every 10 000 GHMC steps and subjected them to Langevin simulation [Eq.~\eqref{BusPar}] at fixed volume using a collision rate of 9.1/ps. We integrated these initial conditions for a total of 4096 steps using a variety of different time steps from 0.25~fs to 7~fs, with the accumulated shadow work after $2^n$ steps stored ($n = 0, 1, \ldots, 12$). The limit of stability was determined by the largest time step that did not generate infinite cumulative work values in 4096 time steps in any sample; we determined the limit to be 2~fs for unconstrained simulations and 6~fs for constrained simulations.

\begin{figure*}[t]
\begin{center}
\begin{tabular}{cc}
\includegraphics[width=\columnwidth]{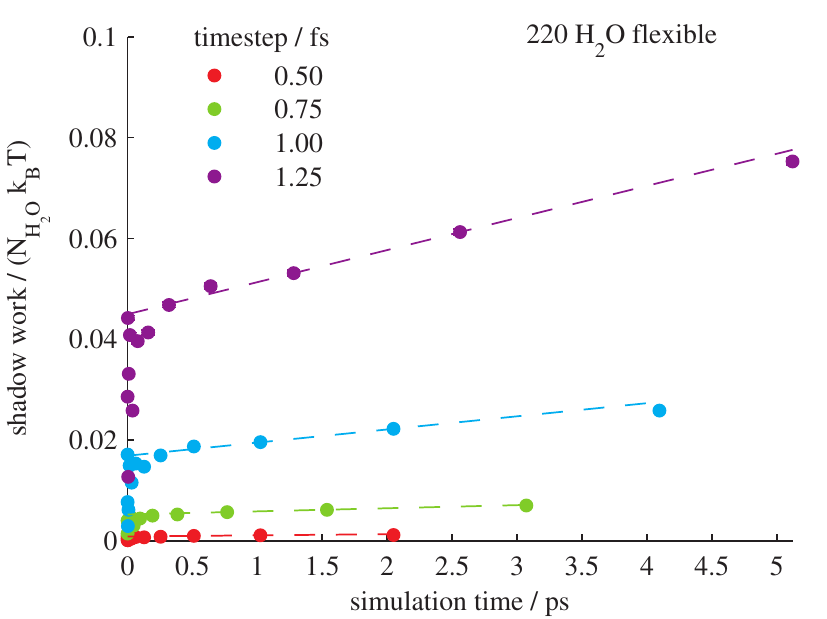} &
\includegraphics[width=\columnwidth]{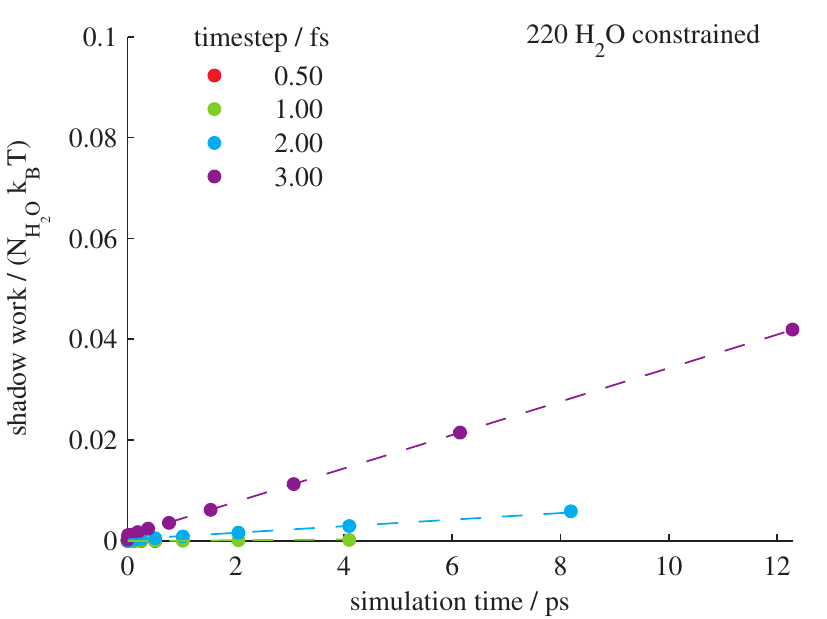} \\ \\ \\
\includegraphics[width=\columnwidth]{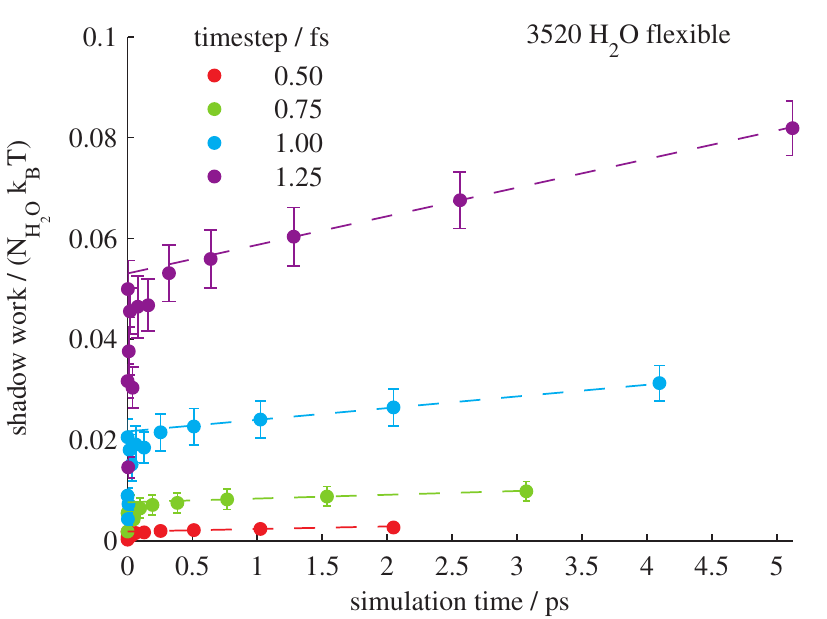} & 
\includegraphics[width=\columnwidth]{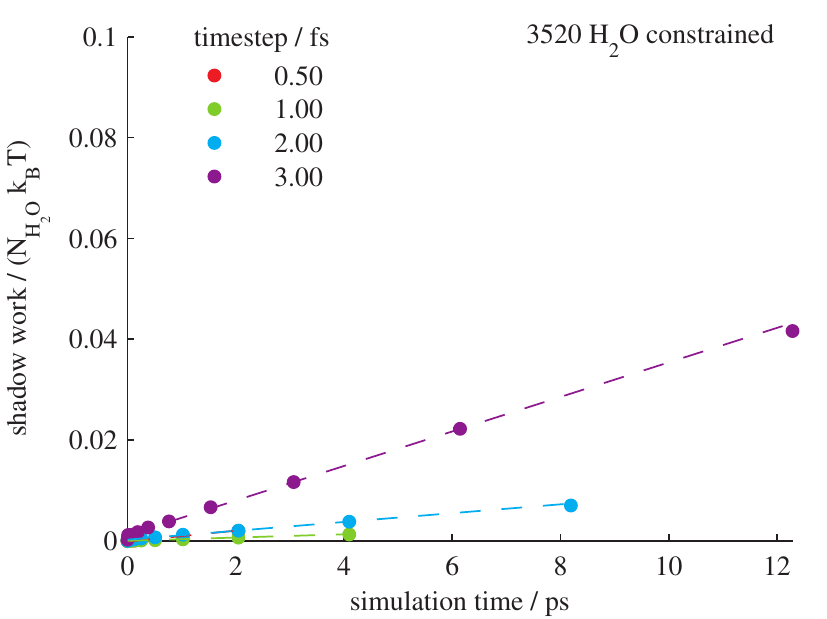}
\end{tabular}
\caption{Convergence to steady state of Langevin simulations with a time-independent Hamiltonian. Shadow work accumulates at a steady rate after $M = 1024$ steps. Each dashed line connects work values at 1024 and 2048 steps. According to Eq.~\eqref{equ:FneqEst}, the nonequilibrium free energy is estimated as half the $y$ intercept of the dotted line. Left column: unconstrained simulations; right column: constrained simulations. Top row: 220 water molecules; bottom row: 3520 water molecules. Each simulation ran for 4096 steps. Error bars denote 95\% confidence intervals.}
\label{fig:converge}
\end{center}
\end{figure*}

To estimate, using Eq.~\eqref{equ:FneqApp}, the nonequilibrium free energy of the steady-state ensemble sampled by discrete Langevin integration, we used the average accumulated shadow work after $M$ steps as the work to switch into steady state, while we used the average dissipated power in the next M steps as an average steady-state power:
\begin{align}
\label{equ:FneqEst}
\Delta F_{\rm neq} = \tfrac{1}{2} \Big[ \left< \work_{0 \rightarrow M} \right>_{\rm GHMC} - \left<\work_{M \rightarrow 2M}\right>_{\rm GHMC} \Big] \ .
\end{align}
Here, the $\left< \cdot \right>_{\rm GHMC}$ notation denotes averages computed over Langevin simulations initiated from GHMC-sampled initial configurations and momenta. Through analysis of $M = 2^n$ for $n = 0, 1, \ldots, 11$, we found that the steady-state power, and hence the estimated nonequilibrium free energy, converged after $M = 1024$ steps (see Fig.~\ref{fig:converge}), so we used this value for all subsequent analysis.

We estimated the squared uncertainty in the nonequilibrium free energy as
\begin{align}
\label{equ:waterUncertain}
\delta^2 (\Delta F_{\rm neq}) &= \Big[ \var{\work_{0 \rightarrow M}} + \var{\work_{M \rightarrow 2M}} \notag\\
- 2\ &\cov{\work_{0 \rightarrow M}, \work_{M \rightarrow 2M}} \Big] \big/ (4 N_{\rm eff})
\end{align}
where $\var{x}$ and $\cov{x,y}$ denote sample variances and covariances over the measured set of work values, and $N_{\rm eff}$ is the effective number of uncorrelated samples after accounting for the statistical inefficiencies by autocorrelation analysis of sequentially-sampled trajectory work values (see Sec. 2.4 of Ref.~\cite{Chodera:2007jb}).

%

\end{document}